\documentclass[prl,aps,twocolumn]{revtex4}
\usepackage{graphicx} 
\usepackage[usenames]{color}
\usepackage{amsmath,amssymb}
\usepackage{gensymb}
\usepackage{natbib}
\usepackage{xcolor}
\def\strutdepth{\dp\strutbox}
\def\nw#1{\strut\vadjust{\kern-\strutdepth\vtop to0pt{\vss\hbox to\hsize
{\hskip\hsize\hskip5pt$\leftarrow$\hss\strut}}}{\em #1}}
\usepackage{rotating}

\usepackage{color}
\definecolor{red}{rgb}{1,0,0}

\begin{document}

\title{Extended $\mu(J)$-rheology for dense suspensions at oscillatory shear flows}

\author{Junhao Dong and Martin Trulsson}
\affiliation{Theoretical Chemistry, Lund University, Sweden}

\begin{abstract}
Recent studies have highlighted that oscillatory and time-dependent shear flows might help increase flowability of dense suspensions. While most focus has been on cross-flows we here study a
simple two-dimensional suspensions where we apply simultaneously oscillatory and stationary shear along the same direction. We first show that the viscosities in this set-up significantly decrease with an increasing magnitude of the oscillations, contrary to previous claims. The decrease can be attributed to the large decrease in the number of contacts and an altered microstructure as the magnitude of the oscillation is increased. As a sub-result we find both an extension to the $\mu(J)$-rheology, a constitutive relationship between the shear stresses and the shear rate, valid for pure oscillatory flows and with a higher shear-jammed packing fraction for suspensions composed of frictional particles compared to steady-shear conditions.
\end{abstract}


\date{\today}
\maketitle

The flow of dense suspensions, even for the simplest systems, show several non-trivial rheological behaviours such as shear-thinning \cite{Colin17}, both continuous \cite{Trulsson12, Ness15} and discontinuous shear-thickening \cite{Seto13, Wyart14, Dong17,Jamali19}, and shear jamming \cite{Olsson07,Bi11}.  In the simplest models of a non-Brownian suspension composed of hard bodies the viscosity $\eta$ diverges as $\eta/\eta_f \sim (\phi_c-\phi)^{-\alpha}$, where $\eta_f$ is the solvent viscosity, $\phi$ the packing fraction, and $\alpha$ a critical exponent usually close to 2 \cite{Boyer11,Heussinger12} and where $\phi_c$ depends on the friction \cite{Silbert10}, shape \cite{Donev04,Azema15,Nagy17,Trulsson18, Silbert18,Marschall19a, Marschall19b}, and interactions \cite{Irani14, Berger15}. 
The shear-thinning and shear-thickening can usually be attributed to a decreased or an increased importance of certain interactions compared to others.
For example, shear-thinning of colloidal suspensions is due to a decreased importance of thermal collisional interactions (vibrations in ``soft cages'') compared to the shear stresses \cite{Trulsson15}.
Shear-thickening can be attributed to an increased importance of hydrodynamics \cite{Wagner09,Jamali19} or as in the case of discontinuous shear-thickening the onset of frictional interactions above a certain stress threshold  \cite{Seto13,Wyart14}.  Shear-thickening can also be driven by inertial effects \cite{Trulsson12, Ness15} or a combination of them \cite{Ness16, Dong20}. \newline
Acoustics \cite{Sehgal19} and oscillatory flows \cite{Lin16, Ness18} have recently been shown to be effective in altering the flowability of suspension, where one has the ability to lower the the viscosity in a controlled manner (\textit{i.e.}~the resistance to flow). Especially cross-flows have turned out to be useful when an increased flowability of (almost) shear jammed dense suspensions is desired and can even under certain circumstances unblock shear jammed suspensions \cite{Ness18}. The current understanding of this decrease in shear viscosity with cross-flows is based on a force chain tilting as a consequence to this secondary oscillatory shear flow \cite{Lin16} and a random organisation \cite{Ness18}, the later a concept closely related to reversibility of oscillatory shear flows for suspensions \cite{Pine04}.
\newline
In this paper we show that also oscillatory flows parallel to an average shear flow decrease a generalised complex viscosity and number of contacts as well as altering the microstructure. We, furthermore, show that the shear-jamming point for a suspension composed of frictional particles shifts to higher packing fractions as large oscillations are applied and approaches a new shear jamming packing fraction with a value just below the shear jamming point for suspensions composed of frictionless particles. For suspensions composed of frictionless particles the point of shear jamming is not shifted but the viscosity is reduced by roughly one order of magnitude.\\
 \newline
We consider suspensions composed of roughly 1000 polydisperse discs  sheared between two rough walls, created by fusing particles together, at constant packing fractions and with a time-dependent shear rate $\dot\gamma=\dot\gamma_0+\dot\gamma_1\cos(\omega t+\delta)$,
where $\dot\gamma_0$ is the average shear-rate, $\dot\gamma_1$ the maximum oscillatory shear-rate, $\omega/2 \pi$ the frequency of oscillatory shear, and $\delta$ a shift in time. Without any loss of generality, we assume from now on $\delta$ to be equal to zero and both $\dot \gamma_1$ and $\dot \gamma_0$ to be positive numbers.
Particles interact via normal and tangential forces (frictional forces) with a Coulomb criteria for sliding where we set the particle friction coefficient $\mu_p$ equal to either 0.4 (frictional) or 0 (frictionless). Particles also interact with the solvent via hydrodynamic drag and torques, both linear in the translational and angular velocity differences. 
For more details of the simulations see \cite{Trulsson17, Dong17}.
\begin{figure}[t!]
\includegraphics[scale=0.71]{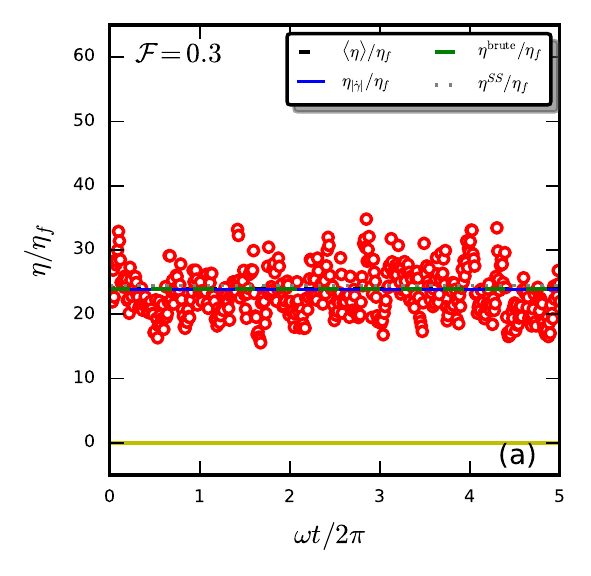}
\includegraphics[scale=0.71]{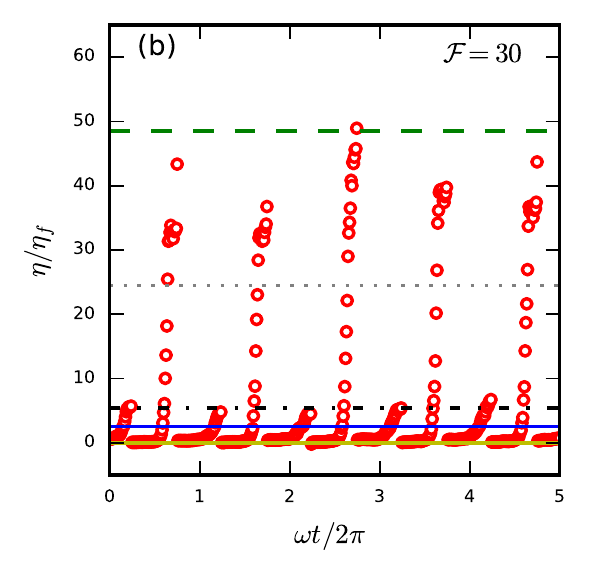}
\caption{Instantaneous relative viscosities at $\phi=0.67$ and $\mu_p=0.4$,  $\mathcal{G}=0.33$ and (a) $\mathcal{F}=0.3$ and (b) $\mathcal{F}=30$. Lines give viscosities estimated through brute force $\eta^{\rm brute}/\eta_f$ (green-dashed), strain-averaged time-average $\eta_{|\dot\gamma|}/\eta_f$ (solid-blue), and normal time-average $\langle \eta \rangle/\eta_f$ (black-dash-dotted); grey dotted line indicates viscosity for steady shear (SS) flow at $\phi=0.67$ (\emph{i.e.}~$\mathcal{F}=0$), yellow solid line indicates zero line. } 
\label{Fig1}
\end{figure}
While the viscosity in a steady shear flow is defined by $\eta^{\rm brute} = \langle \sigma \rangle/ \langle \dot \gamma \rangle$, where $\sigma$ is the shear stress, it becomes obvious that this expression becomes imprecise and possibly invalid at pure oscillatory flows where $\langle \sigma \rangle \to 0$ and $\langle \dot \gamma \rangle \to 0$. For pure oscillatory shear flows ($\dot \gamma_0=0$) one instead relies on the complex viscosity $\eta^*$. Assuming pure viscous response this complex viscosity $\eta^*$ is equal to its viscous part $\eta''$ as  $\eta^*=\eta''=\frac{\int_0^{2 \pi/\omega}\sigma(t) \cos(\omega t)\, dt}{ \dot \gamma_1 \int_0^{2 \pi/\omega} \cos^2(\omega t)\, dt}= G''/\omega$, being essentially a strain weighted quantity, with $G''=\frac{\omega^2}{\pi \dot \gamma_1} \int_0^{2 \pi/\omega}\sigma(t) \cos(\omega t)\, dt$ being the loss modulus \cite{Ewoldt08}.
We generalise this quantity to define strain weighted time-averaged quantities as:
\begin{equation}\label{eq:avg}
\langle A \rangle_{|\dot \gamma|}=\frac{\int_0^{2\pi n /\omega} |A(t)||\dot\gamma(t) | dt}{\int_0^{2\pi n /\omega} |\dot\gamma(t) | dt},
\end{equation}
where $A(t)$ is a time-dependent quantity of interest (\emph{e.g.}~shear stress $\sigma$ or number of contacts $Z$) and an integer $n$ (number of oscillation periods of the time-average). Strain-weighted viscosities can then be calculated as $\eta_{|\dot \gamma|}=\langle \sigma \rangle_{|\dot \gamma|}/\langle \dot \gamma \rangle_{|\dot \gamma|}$, in perfect agreement with the viscous part of the complex viscosity for a pure oscillatory shear flow. 
For a Newtonian suspension with $\sigma = \eta^{SS}(\phi) \dot \gamma$ and an instantaneous response to changes in shear-rate the two expressions, brute and strain-weighted, yield the same viscosity equal to $\eta^{\rm brute}=\eta_{|\dot \gamma|}=\eta^{SS}(\phi)$.
However, for time-dependent flows with delay in or in other way altered shear stress response one has $\eta^{\rm brute} \neq \eta_{|\dot \gamma|}$.
The power dissipated in these flows are time-dependent $\mathcal{D}(t)=\sigma (t) \dot \gamma(t)$. By normalising the average power of dissipation $\langle \mathcal{D}(t) \rangle$ with $\langle \dot \gamma^2(t) \rangle$ we obtain that $\langle \mathcal{D}(t) \rangle/\langle \dot \gamma^2(t) \rangle=\langle \sigma \rangle_{|\dot \gamma[}/\langle \dot \gamma \rangle_{|\dot \gamma|}$, \textit{i.e.}~equivalent to our strain-averaged viscosity.\\
Simulations were typically carried out with either $n=10$ or a total absolut strain of $\int |\dot \gamma | \, dt=10$. Before starting to measure we pre-sheared all sampled for a few oscillatory periods or with a minimum of one absolute strain.
We report both instantaneous time-series of $\eta/\eta_f$ as well as strain-average quantities (according to Eq.~\ref{eq:avg}) of $Z$, $\eta$, and later on also $\phi$, $J$ (viscous number), and $\mu$ (stress ratio). Viscosities are compared to the steady-shear viscosities $\eta^{SS}$ found for the same packing fractions of the system.
\newline
For this simple system we can identify three dimensionless parameters: $\mathcal{F}=\dot\gamma_1/\dot\gamma_0$, $\mathcal{G}=\dot\gamma_1/\omega$, and $\phi$.
 $\mathcal{G}$ gives here the maximum of strain due to oscillations.
Fig.~1 shows two typical cases of how the viscosities varies with time and two different oscillation amplitudes.
When $\mathcal{F}<1$ (small to moderate shear-rate oscillations compared to the average flow) the viscosity only mildly fluctuates around an average viscosity equal to that of $\eta^{SS}(\phi)$ seen in steady state \cite{Trulsson12}, see Fig.~1(a). However, when $\mathcal{F}>1$ (large shear-rate oscillations compared to the average flow) two distinct and alternating peaks  appear each period separated by zones with almost zero viscosities, see Fig.~1(b).
Ones sees that the measure $\eta^{\rm brute}$ poorly captures an average viscosity and is closer to the peak viscosity. In general $\eta_{|\dot \gamma|}$ performs better in capturing the average viscosity and give better agreements with the time-averaged viscosity $\eta^{\rm time} = \omega \int_0^{2\pi n /\omega} \eta(t) \, dt /(2\pi n)$.
\begin{figure}[t!]
\includegraphics[scale=0.66]{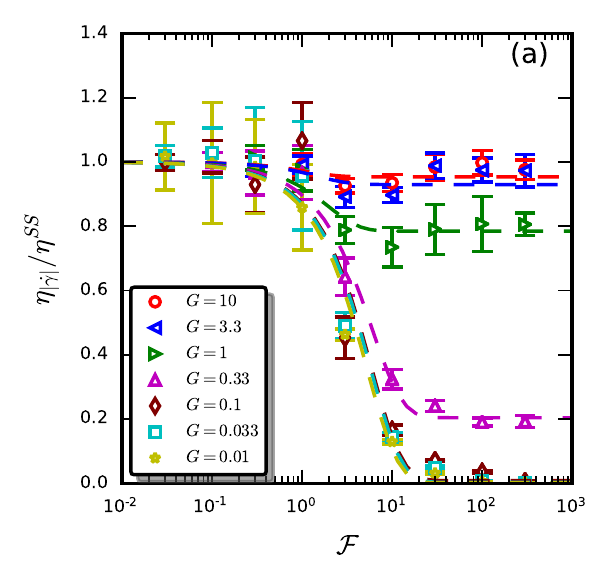}
\includegraphics[scale=0.66]{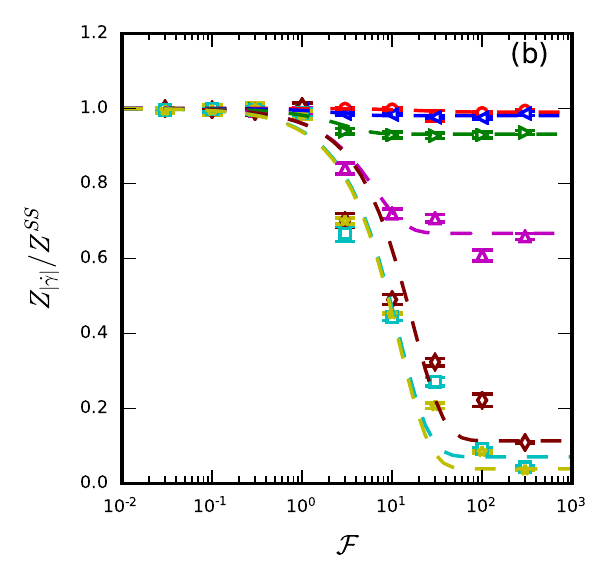}
\includegraphics[scale=0.66]{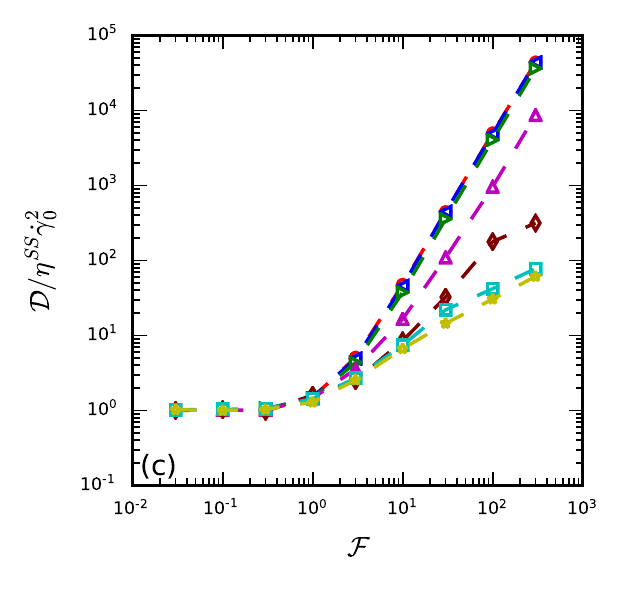}
\includegraphics[scale=0.66]{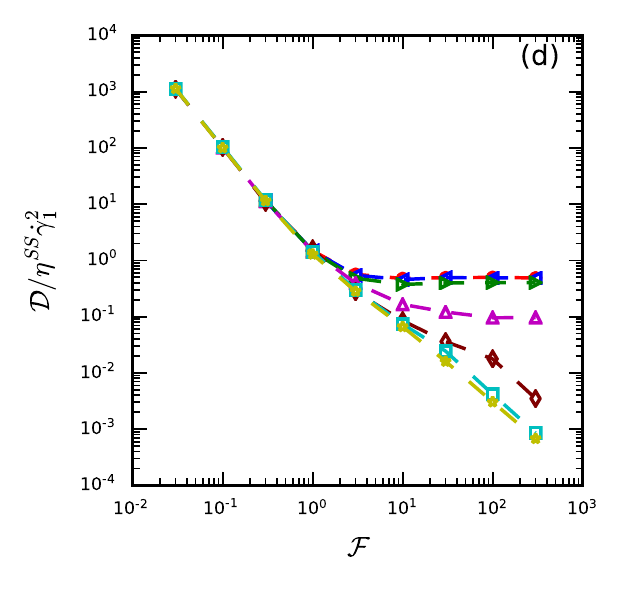}
\caption{(a) Reduced viscosity $\eta_{|\dot\gamma|}/\eta_f$, (b) $Z_{|\dot\gamma|}$, (c) $\mathcal{D}/\dot \gamma_0^2$ and (d) $\mathcal{D}/\dot \gamma_1^2$ as function of $\mathcal{F}$ at various $\mathcal{F}/\mathcal{G}$, $\phi=0.76$ and $\mu_p=0.4$; dashed lines in (a) and (b) are best fits using a hyperbolic tangent function $A_{|\dot\gamma|}/A^{SS}=1-c_1 \tanh (c_2\mathcal{F})$, where $c_1=(1-A_{|\dot \gamma|}^{\mathcal{F}=\infty}/A^{SS})$ and $c_2$ are two free parameters and $A$ are either $\eta_{|\dot \gamma|}/\eta_f$ or $Z_{|\dot \gamma|}$, and dashed lines in (c) and (d) are $\eta_{|\dot \gamma|}/\eta^{SS} (1+0.5 \mathcal{F}^2)$ and $\eta_{|\dot \gamma|}/\eta^{SS} (\frac{1}{\mathcal{F}^2}+0.5)$ with $\eta_{|\dot \gamma|}$ obtained from (a). }
\label{Fig2}
\end{figure}
\begin{figure*}[t!]
\includegraphics[scale=0.90]{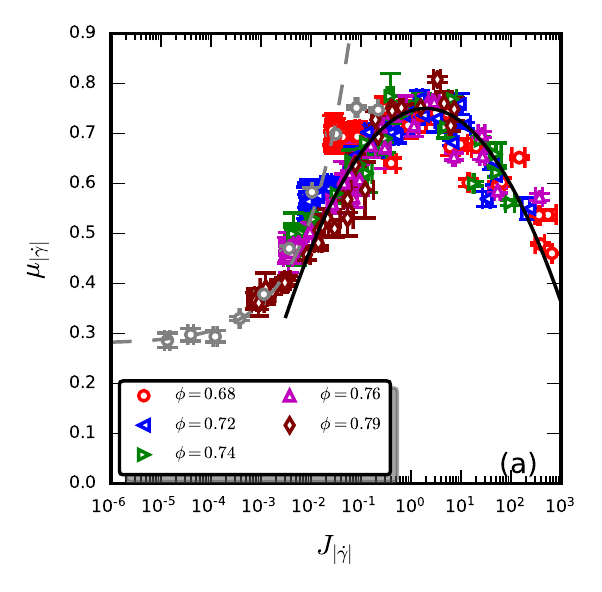}
\includegraphics[scale=0.90]{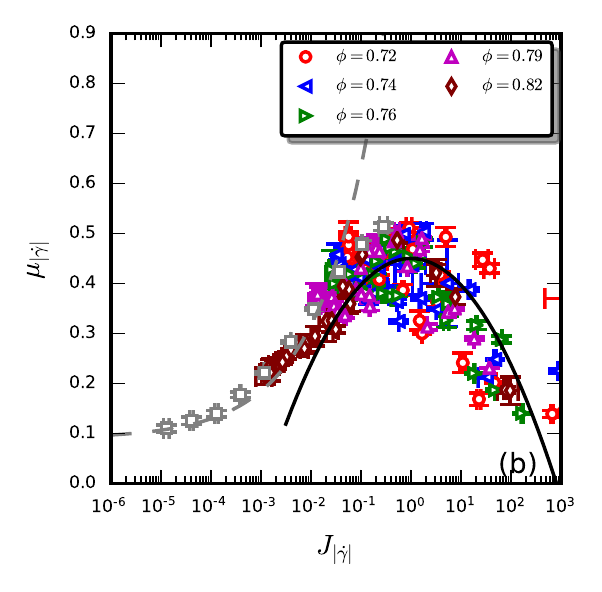}
\includegraphics[scale=0.90]{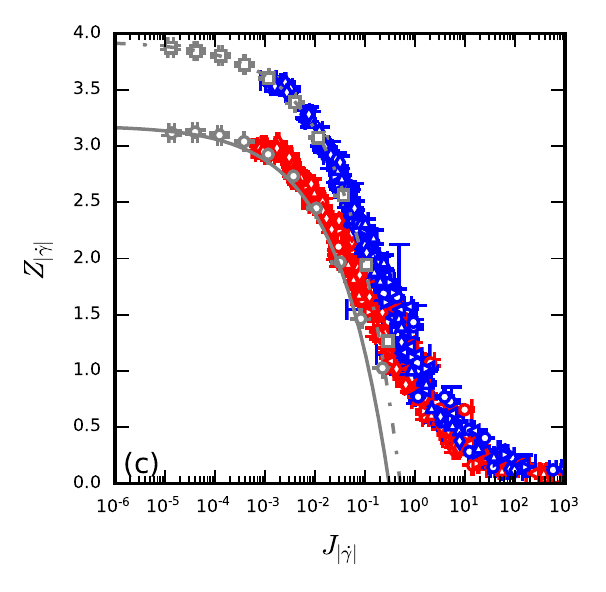}
\caption{Flow curves with extended $\mu(J)$-rheology for (a) $\mu$ for suspensions with frictional particles ($\mu_p=0.4$), (b) $\mu$ for suspensions with frictionless particles ($\mu_p=0$), and (c) $Z$ for both suspensions composed of either frictional (red symbols) or frictionless particles (blue symbols). Grey symbols corresponds to steady-state values \cite{Dong20} for (circles) frictional and (squares) frictionless particles, lines according to \cite{Dong20}. Various symbols and colors in (a) and (b) corresponds to various $\phi$, $\mathcal{F}$, and $\mathcal{G}$ values.} 
\label{Fig3}
\end{figure*}
\begin{figure}[t!]
\includegraphics[scale=0.675]{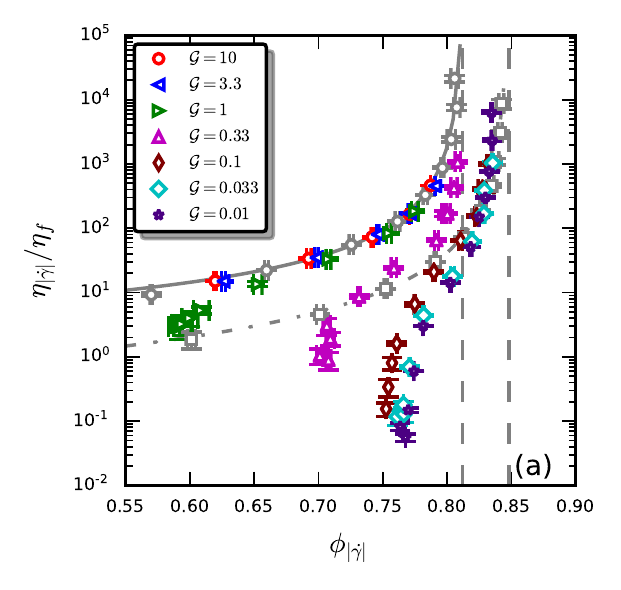}
\includegraphics[scale=0.675]{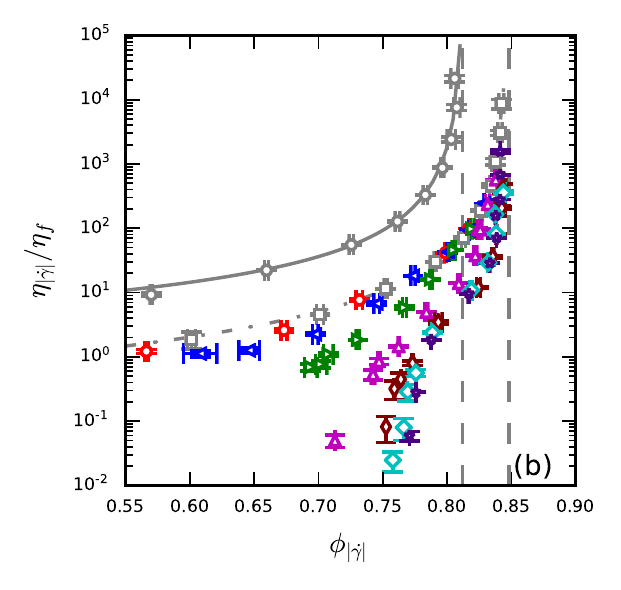}
\caption{Reduced viscosity as function of $\phi$ at various $\mathcal{G}$ and $\mathcal{F}=\infty$ (pure oscillatory flows). Suspension composed of (a) frictional and (b) frictionless particles. Grey symbols correspond to steady-state viscosities for (circles) frictional and (suqare) frictionless. Values are taken from \cite{Dong20}. The grey lines corresponds to best fits according to $\eta(\phi)/\eta_f = a(\phi_c^{SS}-\phi)^{-\alpha}$ for (solid) frictional and (dash-dotted) frictionless particles. Vertical grey dashed lines show the locations of the corresponding shear-jamming points, with  and $\phi_c^{SS,\rm frictional}=0.848\pm 0.002$ and $\phi_c^{SS,\rm frictionless}=0.812\pm 0.002$.} 
\label{Fig4}
\end{figure}
In Fig.~2 we show how (a) the strain-averaged viscosities $\eta_{|\dot\gamma|}/\eta_f$, (b) number of contacts $Z_{|\dot \gamma|}$, dissipation $\mathcal{D}$ normalised by (c) $\dot \gamma_0^2$ or (d) $\dot \gamma_1^2$ vary with $\mathcal{F}$ at various oscillating strains, $\mathcal{G}$, at $\phi=0.76$. For all $\mathcal{G}$ the viscosities are close to $\eta^{SS}(\phi)$ as soon as $\mathcal{F}<1$ in agreement with previously reported results \cite{Ness18}. At $\mathcal{F}>1$ the viscosities decrease, moderately for large $\mathcal{G}$ and substantially for small strain amplitude (\emph{i.e.}~small $\mathcal{G}$). The same trends are seen in the number of contacts, highlighting that the former is a consequence of the later. In particular,  we see that the viscosity and number of contacts both decrease to zero at high values of $\mathcal{F}$ and low values of $\mathcal{G}$. We interpret this as being in meta-reversible (finite $\mathcal{F}$) or reversible states ($\mathcal{F}=\infty$) as has previously been observed for pure oscillatory shear flows of suspensions \cite{Pine04,During09,Jeanneret14}. Meta-reversible as the states are only reversible for a certain time and will eventually be broken due to the directed average flow .
\newline 
Our results are hence in line what has been previously found for cross-flows \cite{Ness18}, but here we clearly show that even parallel oscillatory flows are sufficient for increased flowability (\emph{i.e.}~lowered viscosity).
 This puts new doubt on the explanation about the tilting of the force chains, as there is no possibility for this in 2D. Instead the decrease of contacts, and hence all types of force chains, seems to be the main explanation. As a note, using the strain-weighted measure for cross-flows would leave previous results unaltered as they were measured along a direction where the shear-rate was constant in time.\newline
As discussed by Ness \textit{et al.}~\cite{Ness18} the dissipation normalised by the average strain might be a more interesting quantity, especially for several industrial applications where one wants the lower the energy consumption per strain.  Looking at this quantity, see Fig.~2(c), one finds similar findings as seen for cross-flows. At low $\mathcal{F}$ the dissipation remains equal to that at steady-state shear. However, as soon as $\mathcal{F}>1$ the dissipation increases. In other words, one finds that $\mathcal{D}/ \dot \gamma_0^2 = \eta_{|\dot \gamma|} (1+0.5 \mathcal{F}^2)$ and $\mathcal{D}/ \dot \gamma_1^2 = \eta_{|\dot \gamma|} (1/\mathcal{F}^2+0.5)$. The later might be a more interesting quantity for flows which are almost purely oscillatory shear flows. 
Notice that $\eta_{|\dot \gamma|}$ is itself dependent on $\mathcal{F}$, as well as $\mathcal{G}$ and $\phi$.
While we get a monotonic increase for the dissipation both from (c) a pure steady shear flow and (d) a pure oscillatory shear flow, cross-flows seems to show a modest non-monotonicity \cite{Ness18}. This could possibly be attributed to that we neglected hydrodynamic interactions between pairs of particles or indeed that having the oscillations perpendicular to the average flow is slightly more beneficial in regards to lowering the dissipation per strain.
\begin{figure*}[t!]
\includegraphics[scale=0.5]{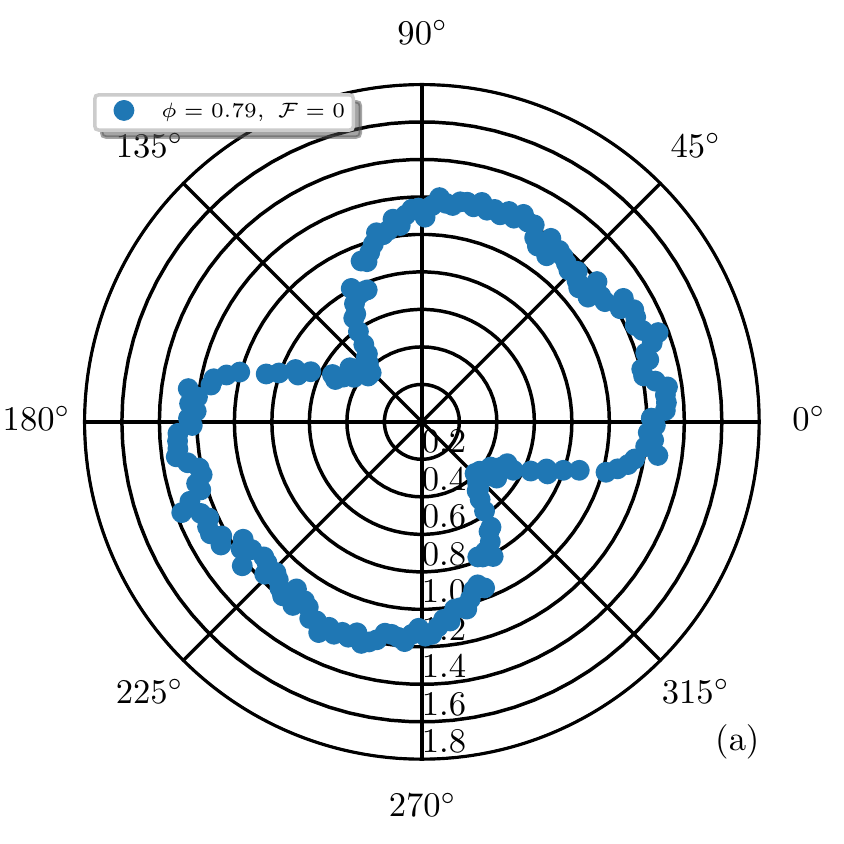}
\includegraphics[scale=0.5]{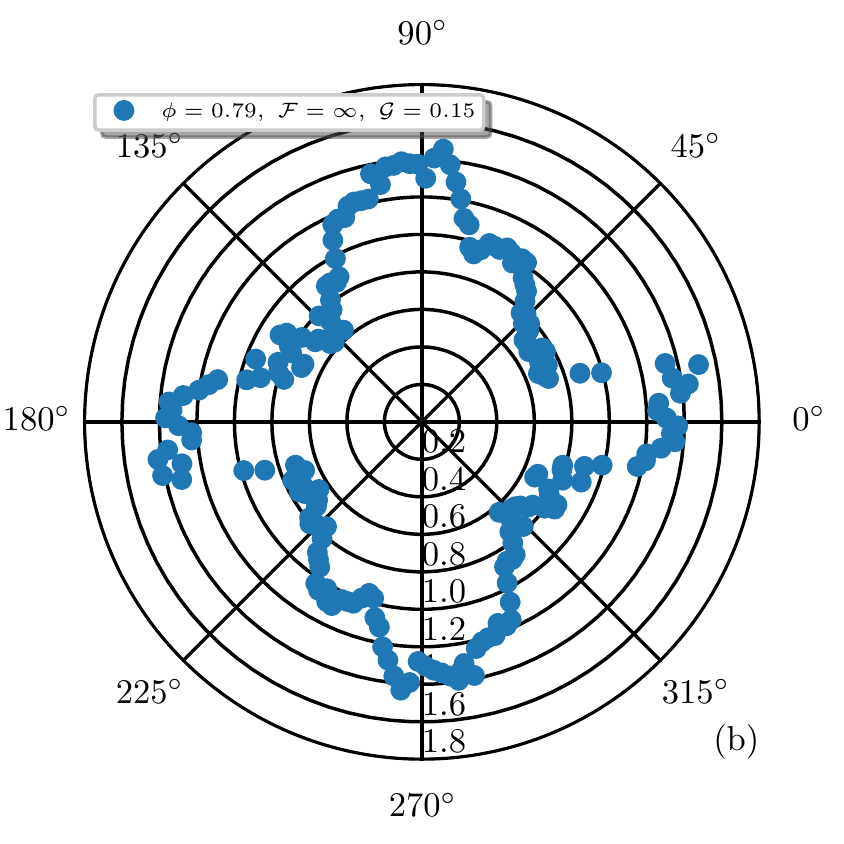}
\includegraphics[scale=0.5]{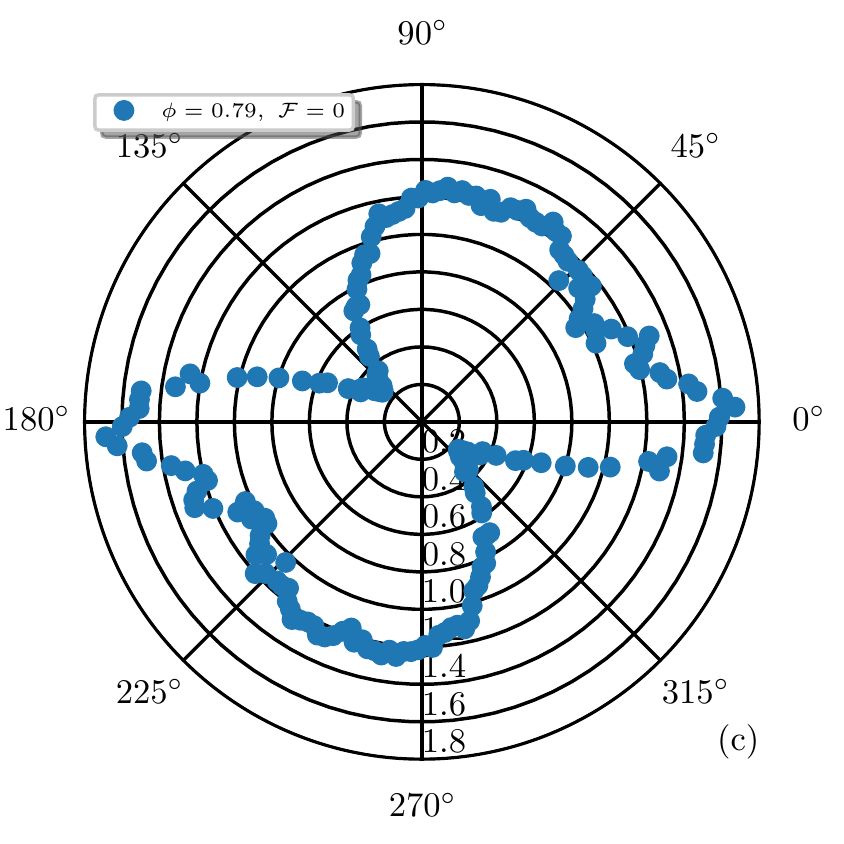}
\caption{Normalised polar contact probability functions of the particles at $\phi=0.79$. (a) steady shear of frictional particles with $\eta_{|\dot\gamma|}/\eta_f=451$, (b) oscillatory shear with $\eta_{|\dot\gamma|}/\eta_f=29$, and (c) steady shear of frictionless particles with $\eta_{|\dot\gamma|}/\eta_f=26$.}
\label{Fig5}
\end{figure*}
An important finding from above is that $\eta_{|\dot \gamma|, \mathcal{F}=\infty} \leq \eta_{|\dot \gamma|, \mathcal{F}=0}=\eta^{SS}(\phi)$ for the same $\phi$, where viscosities for oscillatory flows and steady state differ for small to moderate oscillatory strains ($\mathcal{G}<3.3$), hence do not respect the Cox-Merz rule, and are approximately equal otherwise.
Interestingly, there exists a small regime ($1<\mathcal{F}<10$) for which the viscosities are lower both that of steady shear flow and oscillatory shear flow viscosities even for large oscillatory strains.\\
\newline
Suspension rheology can be re-formulated using a viscous number $J=\eta_f \dot \gamma /P $ where $\phi(J)$, $\mu(J)$, and $Z(J)$ all are constitutive relationships and functions of only $J$ in the hard-body limit \cite{Boyer11, Trulsson12}, where $P$ is the pressure and $\mu=\sigma/P$ the stress ratio. We here show that one can expand this formulation to oscillatory flows by using a strain weighted viscous number $J_{|\dot \gamma|}= \eta_f \langle \dot \gamma /P\rangle_{|\dot \gamma|}$, where $P$ is measured in the center of the cell.
By doing so we find a collapse of $\mu_{|\dot \gamma|}(J_{|\dot \gamma|})$ and $Z_{|\dot \gamma|}(J_{|\dot \gamma|})$ as soon as $\mathcal{F}\gg1$ for a large parameter space of $\phi$ and $\mathcal{G}$, see Fig.~3, with a cross-over between steady-state and oscillatory dominated flows around $\mathcal{F}\sim 1$.
Data points for which $\mathcal{F}\ll1$ are all well captured by the original $\mu(J)$-rheology whereas data points for which $\mathcal{F}\gg1$, and with $J_{|\dot \gamma|} \gg J^{SS}(\phi)$ 
, are better described by the empirical relation 
$\mu_{|\dot \gamma|} \simeq \mu^{\rm max} -\kappa \Big(\ln(J_{|\dot \gamma|})-\ln(J_{|\dot \gamma|,0})\Big)^2$, with $\mu^{\rm max} \simeq 0.75$, $J_{|\dot \gamma|,0} \simeq 2$, and $\kappa=0.01$, shown in Fig.~3(a) as a black solid line.
A similar trend can be found for suspensions composed of frictionless particles, even though the collapse is slightly worse, as seen in Fig.~3(b). Here the black solid line is the same constitutive law as in Fig.~3(a) but with $\mu^{\rm max} \simeq 0.45$, $J_{|\dot \gamma|,0} \simeq 1$, and $\kappa=0.01$. The collapse works slightly better considering $Z_{|\dot \gamma|}$ plotted against $J_{|\dot \gamma|}$, see Fig.~3(c). \newline
We now explore if one can cross the shear-jamming packing fraction by having shear flow oscillations. We achieve this by doing pressure imposed simulations rather than constant volume. This replaces the control parameter $\phi$ by $J_{|\dot \gamma|}$. Indeed, as observed by cross-flow oscillations we find that the shear-jamming point shifts to higher packing fraction compared to steady-state for flows composed of frictional particles. Fig.~4 shows the viscosity in the limiting case where we have pure oscillatory flows (\emph{i.e.}~$\mathcal{F}=\infty$) at various fixed oscillatory strains. 
The viscosities follow the steady-state viscosities for the corresponding system as soon as the oscillatory strain $\mathcal{G}$ is large. As the oscillatory strain is lowered the 
viscosities of suspensions composed of frictional particles start to follow the viscosity branch corresponding to suspensions composed of frictionless particles at steady shear state in a narrow regime close to the frictionless shear jamming point. This new and partially unexplored jamming point for oscillatory flows which, as we denote $\phi_c^{OS}$ as compared to the steady state jamming point $\phi_c^{SS}$, will hence be dependent on  $\mathcal{G}$ with $\phi_c^{OS, \rm frictional} \to \phi_c^{SS, \rm frictional}$ as $\mathcal{G} \to \infty$ and $\phi_c^{OS, \rm frictional} \to 0.835 < \phi_c^{SS, \rm frictionless}$ as $\mathcal{G} \to 0$.  The transition is found to occur around $\mathcal{G}=0.1$, similar to what is found in experiment \cite{Lin16}.
We do not yet know if this transition is continuous or discontinuous.
For suspensions composed of frictionless particles the shear-jamming point seems not to shift for pure oscillatory flow at imposed pressure (\emph{i.e.}~$\phi_c^{OS,\rm frictionless}\approx \phi_c^{SS,\rm frictionless}$). The viscosities do, however, decrease by roughly one order of magnitude. Our results are in agreement with previous findings for oscillatory cross-flows \cite{Ness18} for low $\mathcal{G}$ with the exception that we also find an increased flowability also for frictionless particles at oscillatory flows and low $\mathcal{G}$. \\ 
As a final test we study the microstructure of steady-state shear (frictional), oscillatory shear flow (frictional), and steady-state shear (frictionless) samples at constant packing fraction $\phi=0.79$. Fig.~5 shows how the contact distribution changes from a two-fold rotational symmetry for steady shear flows to having a four-fold symmetry for oscillatory shear flows. Even if the pure oscillatory flow of the frictional particles has a viscosity similar to that of a frictionless sample at the same packing fraction the two microstructures are very different indicating that the ``mechanism'' for jamming is not the same in the two approaches and the collapse of the oscillatory viscosities at low strains onto the steady-state frictionless branch is probably fortuitous. \\ 
\newline
In this paper we have shown that (i) oscillatory shear flows parallel to the average flow leads to a decrease in viscosities, (ii) the $\mu(J)$-rheology can be extended to oscillatory shear flows, and (iii) the oscillatory shear-jamming packing fraction is unaltered compared to steady shear for frictionless particles but shifted upwards for frictional particles.
Our understanding of why this shift occurs in the frictional but not in the frictionless case is that for oscillatory flows with small strains the tangential springs do not have time to get enough strained to mechanically stabilise the suspensions. Frictionless particles lacks this possibility altogether and are hence unaffected by this effect. Hence, this opens up for the alternative strategy of using shear oscillations along an average shear to unblock shear jammed dense suspensions of frictional particles.
\newline
It would be fruitful to expand the $\mu(J)$-rheology to granular rheology in line with ref.~\cite{Ishima19} and using the inertial number $I$ instead of $J$, to explore the role of nonlocal rheology \cite{Rojas15}, and study linear combinations shear oscillations perpendicular to each other with or without an average shear flow.


\begin{thebibliography}{}
\bibitem{Colin17}
%
G Chatt\'e, J. Comtet, A. Nigu\`es,  L. Bocquet, A. Siria, G. Ducouret, F. Lequeux, N. Lenoir, G. Ovarlez, and A. Colin. \textit{Shear thinning in non-Brownian suspensions}. Soft Matter \textbf{14}, 879-893 (2018).

\bibitem{Trulsson12}
M. Trulsson, B. Andreotti and P. Claudin. \textit{Transition from the viscous to inertial regime in dense suspensions}. Phys. Rev. Lett. \textbf{109}, 118305 (2012).

\bibitem{Ness15}
C. Ness and J.Sun. \textit{Flow regime transitions in dense non-Brownian suspensions: Rheology, microstructural characterization, and constitutive modeling}. Phys. Rev. E \textbf{91}, 012201 (2015). 

\bibitem{Seto13}
R. Seto, R. Mari, J.~F. Morris, and M.~M. Denn. \textit{Discontinuous shear thickening of frictional hard-sphere suspensions}. Phys. Rev. Lett. \textbf{111}, 218301 (2013).

\bibitem{Wyart14}
M. Wyart and M.~E. Cates. \textit{Discontinuous Shear Thickening without Inertia in Dense Non-Brownian Suspensions}. Phys. Rev. Lett. \textbf{112}, 098302 (2014). 

\bibitem{Jamali19}
S. Jamali and J.~F. Brady.  \textit{Alternative Frictional Model for Discontinuous Shear Thickening of Dense Suspensions: Hydrodynamics}. Phys. Rev. Lett. \textbf{123}, 138002 (2019). 

\bibitem{Dong17}
J. Dong and M. Trulsson. \textit{Analog of discontinuous shear thickening flows under confining pressure}. Phys. Rev. Fluids \textbf{2}, 081301(R) (2017).

\bibitem{Bi11}

D. Bi, J. Zhang, B. Chakraborty, and R.~P. Behringer. \textit{Jamming by shear}. Nature \textbf{480}, 355?358 (2011).

\bibitem{Olsson07}
P. Olsson and S. Teitel. \textit{Critical Scaling of Shear Viscosity at the Jamming Transition}. Phys. Rev. Lett. \textbf{99}, 178001 (2007).

\bibitem{Heussinger12}
B. Andreotti, J.-L. Barrat, and C. Heussinger. \textit{hear flow of non-Brownian suspensions close to jamming}. Phys. Rev. Lett. \textbf{109}, 105901 (2012). 

\bibitem{Boyer11}
F. Boyer, E. Guazzelli and O. Pouliquen, \textit{Unifying Suspension and Granular Rheology}. Phys. Rev. Lett. \textbf{107}, 188301 (2011)

\bibitem{Silbert10}
L.~E. Silbert. \textit{Jamming of frictional spheres and random loose packing}. Soft Matter \textbf{6}, 2918-2924 (2010). 

\bibitem{Donev04}
A. Donev, I. Cisse, D. Sachs, E. A. Variano, F.~H. Stillinger, R. Connelly, S. Torquato, and P.~M. Chaikin. \textit{Improving the Density of Jammed Disordered Packings Using Ellipsoids}. Science \textbf{303}, 990-993 (2004).

\bibitem{Azema15}
 E. Az\'ema, F. Radjai, and J.-N. Roux. \textit{Internal friction and absence of dilatancy of packings of frictionless polygons}. Phys. Rev. E \textbf{91}, 010202(R) (2015).

\bibitem{Nagy17}
D.~B. Nagy, P. Claudin, T. B\"orzs\"onyi, and E. Somfai. \textit{Rheology of dense granular flows for elongated particles}. Phys. Rev. E \textbf{96}, 062903 (2017). 

\bibitem{Trulsson18}
M. Trulsson. \textit{Rheology and shear jamming of frictional ellipses}. J. Fluid Mech. \textbf{849}, 718-740 (2018).

\bibitem{Marschall19a}
T. Marschall, Y.-E. Keta, P. Olsson, and S. Teitel. \textit{Orientational ordering in athermally sheared, aspherical, frictionless particles}. Phys. Rev. Lett. \textbf{122}, 188002 (2019).

\bibitem{Marschall19b}
T.~A. Marschall and S. Teitel. \textit{Shear-driven flow of athermal, frictionless, spherocylinder suspensions in two dimensions: Stress, jamming, and contacts}. Phys. Rev. E \textbf{100} (3), 032906 (2019).

\bibitem{Silbert18}
K.~M. Salerno, D.~S. Bolintineanu, G.~S. Grest, J.~B. Lechman, S.~J. Plimpton, I. Srivastava, and L.~E. Silbert. \textit{Effect of shape and friction on the packing and flow of granular materials}. Phys. Rev. E \textbf{98}, 050901(R) (2018).

\bibitem{Irani14}
E. Irani, P. Chaudhuri, and C. Heussinger. \textit{Impact of attractive interactions on the rheology of dense athermal particles}.Phys. Rev. Lett. \textbf{112}, 188303 (2014).

\bibitem{Berger15}
N. Berger, E. Az\'ema, J.-F. Douce, and F. Radjai. \textit{Scaling behaviour of cohesive granular flows}. Europhys. Lett. \textbf{112}, 64004 (2015).

\bibitem{Trulsson15}
M. Trulsson, M. Bouzid, J. Kurchan, E. Clément, P. Claudin and B. Andreotti. \textit{Athermal analogue of sheared dense Brownian suspensions}. Europhys. Lett. \textbf{111}, 18001 (2015).

\bibitem{Wagner09}
N.~J. Wagner and J.~F. Brady. \textit{Shear thickening in colloidal dispersions} Phys. Today \textbf{62}, No. 10, 27 (2009).

\bibitem{Lin16}
N.~Y.~C. Lin, C. Ness, M.~E. Cates, J.~Sun, and I.~Cohen, \textit{Tunable shear thickening in suspensions}. Proc. Natl. Acad. Sci. U.S.A. \textbf{113}, 10774 (2016).

\bibitem{Ness16}
C. Ness and J. Sun. \textit{Shear thickening regimes of dense non-Brownian suspensions}. Soft Matter \textbf{12} (3), 914-924 (2016).
 
\bibitem{Dong20}
J. Dong and M. Trulsson. \textit{Unifying viscous and inertial regimes of discontinuous shear thickening suspensions.} J. Rheology \textbf{64} (2), 255-266 (2020).
 
\bibitem{Sehgal19}
P. Sehgal, M. Ramaswamy, I. Cohen, B.~J. Kirby. \textit{Using Acoustic Perturbations to Dynamically Tune Shear Thickening in Colloidal Suspensions}. Phys. Rev. Lett. \textbf{123}, 128001 (2019).
 
\bibitem{Ness18} 
C. Ness, R. Mari, and M.~E. Cates. \textit{Shaken and stirred: Random organization reduces viscosity and dissipation in granular suspensions}. Science Advances \textbf{4} (3), eaar3296 (2018).
 
\bibitem{Pine04}
D.~J. Pine, J.~P. Gollub, J.~F. Brady and A.~M. Leshansky. \textit{Chaos and threshold for irreversibility in sheared suspensions}. Nature \textbf{438}, 997-1000 (2005).
 
\bibitem{Trulsson17}
M. Trulsson, E. DeGiuli, and M. Wyart. \textit{Effect of friction on dense suspension flows of hard particles}. Physical Review E \textbf{95} (1), 012605 (2017).
 
\bibitem{Ewoldt08}
R. H. Ewoldt, A. E. Hosoi, and G. H. McKinley, \textit{J. Rheol.} \textbf{52}, 1427 (2008).

\bibitem{During09}
G. D\"uring, D. Bartolo, and J. Kurchan. \textit{Irreversibility and self-organization in hydrodynamic echo experiments}.Phys. Rev. E \textbf{79}, 030101(R) (2009).

\bibitem{Jeanneret14}
R. Jeanneret and D. Bartolo. \textit{Geometrically protected reversibility in hydrodynamic Loschmidt-echo experiments}. Nat. Commun. \textbf{5}, 3474 (2014).

\bibitem{Ishima19}
D. Ishima and H. Hayakawa, \textit{Dilatancy and compaction in a pressure control granular system under an oscillatory shear}. arXiv: 1902.04759 (2019).

\bibitem{Rojas15}
E. Rojas, M. Trulsson, B. Andreotti, E. Cl\'ement, R. Soto. \textit{Relaxation processes after instantaneous shear-rate reversal in a dense granular flow}. Europhys. Lett. \textbf{109} (6), 64002 (2015).

  
  



\end{thebibliography}
\end{document}